\documentclass[aps,prl,twocolumn,superscriptaddress]{revtex4-1}
\usepackage[colorlinks=true, pdfstartview=FitV, linkcolor=red, citecolor=blue, urlcolor=blue, pdftitle={},pdfauthor={Tomoya Hayata},pdfsubject={}, pdfkeywords={}]{hyperref}
\usepackage{bm,latexsym,amssymb,amsmath,mathrsfs}
\usepackage{graphicx}
\usepackage{color}
\usepackage{times,setspace}
\newcommand{\hmu}{\hat{\mu}}

\begin{document}
\title{Quantum Monte Carlo simulation of intervortex potential in superconductors}
\author{Arata Yamamoto}
\affiliation{Department of Physics, The University of Tokyo, Tokyo 113-0031, Japan}

\begin{abstract}
We investigate the interaction potential of superconducting vortices at the full quantum level.
We formulate the interaction potential in a constrained path integral and calculate it by the quantum Monte Carlo simulation.
The vortex-vortex potential is attractive (type-I), repulsive (type-II), and flat (critical) depending on a coupling constant.
The vortex-antivortex potential also depends on the coupling constant at long range but is always attractive at short range.
\end{abstract}

\maketitle

\paragraph{Introduction.}

Quantum vortices are the macroscopic manifestation of topology in quantum theory.
The vortex in superconductors is one of the most accessible topological defects in laboratory experiments \cite{Abrikosov:1956sx,ESSMANN1967526,PhysRevLett.62.214}.
As a common property of the topological defects, they have topological quantum number, which is stable against perturbation.
Owing to this property, the topological defects behave like quasi-particles with conserved charge.
We can consider the same physical problems as elementary particles with electric charge, say, their interaction, scattering, and bound state.

Let us define the intervortex potential as the interaction energy of two vortices as a function of intervortex distance.
The intervortex potential was studied at the classical level for a long time \cite{PhysRevB.3.3821,PhysRevB.19.4486,PhysRevD.51.1842,Speight:1996px,PhysRevB.65.224504,2011PhRvB..83e4516C}.
The classical analysis however misses quantum fluctuation.
Furthermore, the analysis often relies on assumption or approximation.
The classical analysis cannot be directly generalized to the quantum one.
Since vortices are not explicit degrees of freedom in the Lagrangian, the potential cannot be written by the expectation value of a local operator.
The potential must be redefined on the basis of quantum theory.
Even if the definition is given, the calculation is difficult because vortices are stringy topological defects, which are non-local and non-perturbative collective excitations.
Hence, we need quantum, ab-initio, and non-perturbative analysis to reach the ultimate goal.

Superconductors are classified into type I and type II.
In a type-I superconductor, vortices attract each other.
Multi-vortices merge into a single vortex.
In a type-II superconductor, vortices repel each other. 
There is stable multi-vortex configuration, which is known as the Abrikosov vortex lattice.
The intervortex potential can be used to identify type I and type II, just like an order parameter of a phase transition.
Moreover, the potential tells us more quantitative information of the interaction.
In general, the potential is not a monotonic function of the intervortex distance.
The derivative of the potential is equivalent to whether the interaction is attractive or repulsive.
It depends not only on the type but also on the distance.

In this work, we formulate the full quantum version of the intervortex potential in the path-integral formalism.
We adopt the quantum Monte Carlo simulation to calculate it exactly.
In the previous quantum Monte Carlo and related studies, the intervortex interaction was estimated from the response to external magnetic fields or rotation \cite{Kajantie:1999ih,Mo:2002zz,Hayata:2014kra}, but the intervortex distance was not strictly fixed.
There is another theoretical framework to calculate the potential, that is, a surface operator in dual gauge theory \cite{Hayata:2019rem}, but the correspondence to the original theory is not trivial.

\paragraph{Abelian Higgs model.}

We consider the 3-dimensional Abelian Higgs model in the imaginary-time formalism.
The model consists of the Abelian gauge field $A_\mu$ and the Higgs field $\phi$.
The theory is quantized by the path integral $Z_0 = \int DA D\phi D\phi^* \ e^{-S}$ with the action $S = \int d^3xd\tau \mathcal{L}$.
The $\tau$ direction is periodic and its length $L_\tau$ is equal to the inverse temperature $1/T$.
The path integral can be exactly calculated in the formulation of lattice gauge theory.
In the lattice gauge theory, the path integral is discretized on the (3+1)-dimensional hypercubic lattice.
The Lagrangian density of the lattice Abelian Higgs model is
\begin{equation}
\label{eqL}
\begin{split}
\mathcal{L}(x)
=& \ \frac{\beta}{2} \sum_{\mu,\nu} \{ 1 - \cos F_{\mu\nu}(x) \}
\\
&+ \sum_{\mu} \left| \phi(x+\hat{\mu}) - e^{iA_\mu(x)} \phi(x) \right|^2 
\\
& + \lambda \{ \phi^*(x)\phi(x) - v^2 \}^2
\end{split}
\end{equation}
with the gauge field strength
\begin{equation}
F_{\mu\nu}(x)=A_\mu(x)+A_\nu(x+\hat\mu)-A_\mu(x+\hat\nu)-A_\nu(x)
.
\end{equation}
A shorthand notation is used for the arguments; $x$ stands for the (3+1)-dimensional lattice site $(x,y,z,\tau)$ and $\hat \mu$ stands for the unit lattice vector in the $\mu=x,y,z$ and $\tau$ directions.
All dimensional quantities are scaled by lattice spacing in these and the following equations.

\begin{figure}[t]
\begin{center}
\begin{minipage}{0.24\textwidth}
\includegraphics[width=1\textwidth]{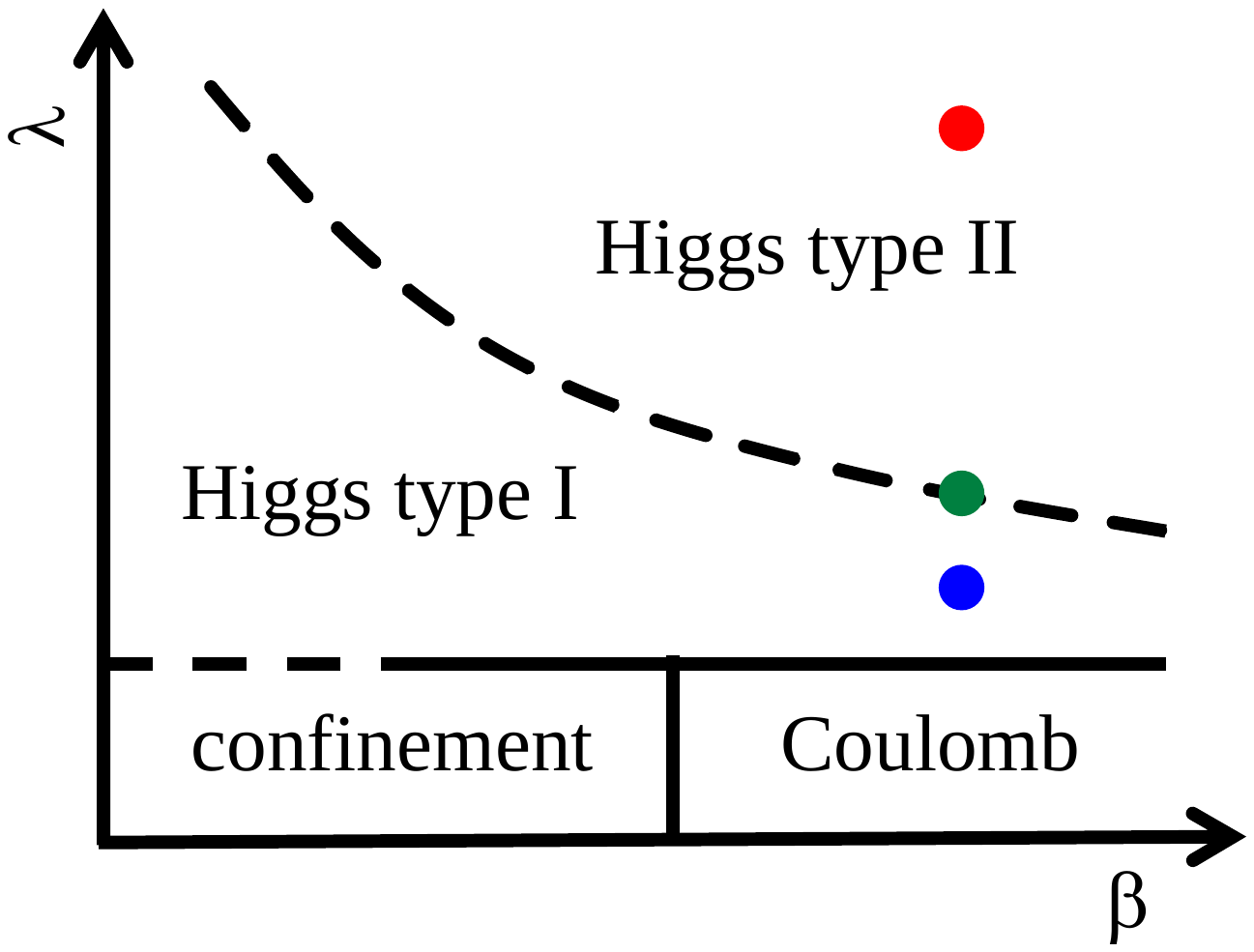}
\end{minipage}
\begin{minipage}{0.23\textwidth}
\includegraphics[width=1\textwidth]{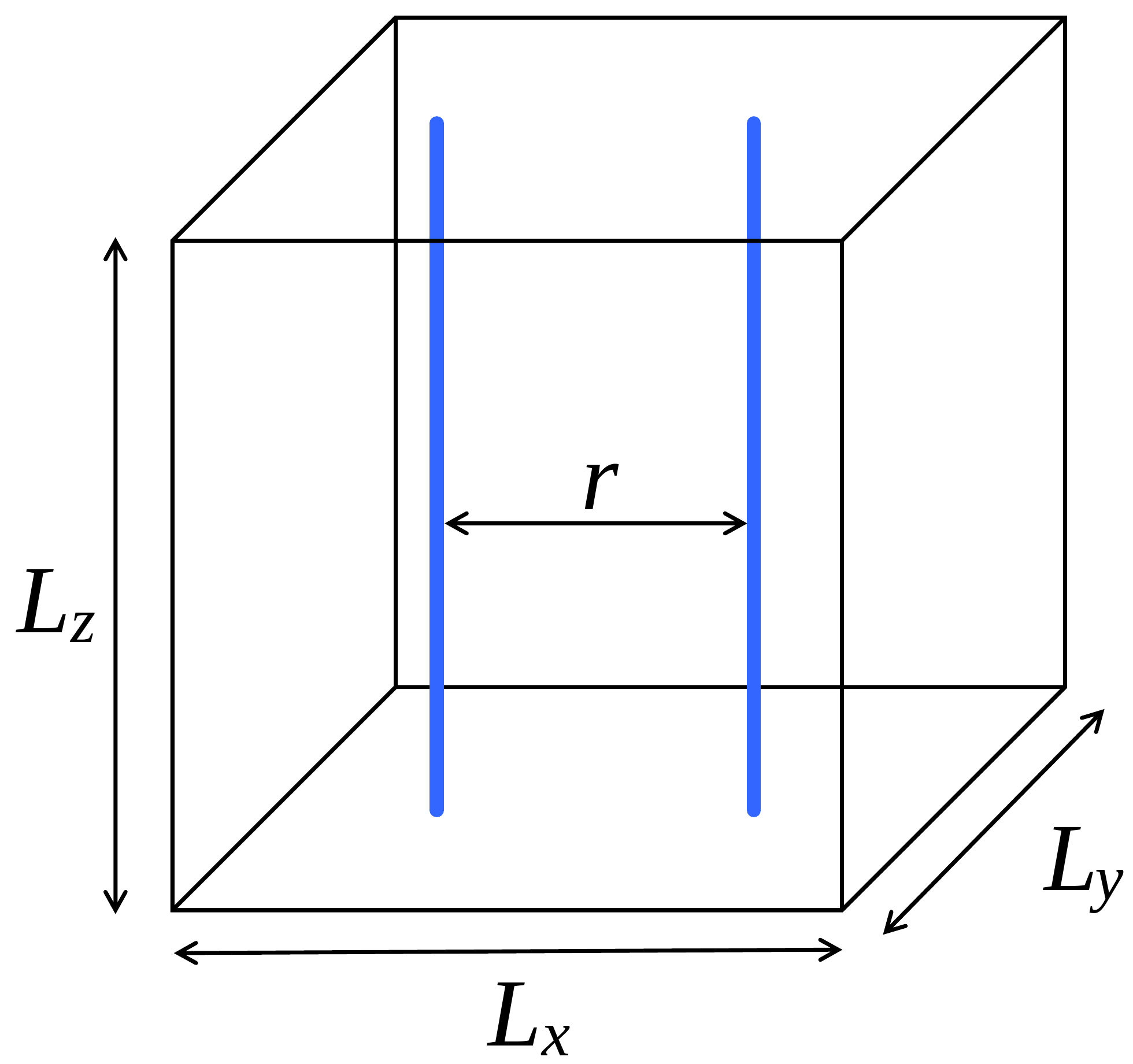}
\end{minipage}
\caption{
\label{fig12}
Left: phase diagram of the lattice Abelian Higgs model.
Three colored dots correspond to the three parameters in Figs.~\ref{figVV} and \ref{figVA}. 
Right: configuration of two vortices.
}
\end{center}
\end{figure}

The model has three parameters: the gauge coupling constant $\beta \equiv 1/e^2$, the scalar self-coupling constant $\lambda$, the condensate parameter $v$.
The schematic phase diagram in the $\beta$-$\lambda$ plane is shown in Fig.~\ref{fig12}.
There are three basic phases: the Coulomb (i.e., normal) phase, the Higgs (i.e., superconducting) phase, and the confinement phase \cite{Fradkin:1978dv}.
The confinement phase is specific to the strongly coupled regime $\beta < \beta_{\rm c} \simeq 1$ and does not exist in the continuum limit.
In the weakly coupled regime $\beta> \beta_{\rm c}$, the Coulomb phase appears in $\lambda<\lambda_{\rm C}$ and the Higgs phase appears in $\lambda>\lambda_{\rm C}$.
They are separated by a first-order phase transition at $\lambda=\lambda_{\rm C}$.
The Higgs phase is further divided into type-I and type-II phases \cite{Damgaard:1988hh}.
The type-I phase appears in small $\beta$ and $\lambda$ and the type-II phase appears in large $\beta$ and $\lambda$.

\paragraph{Intervortex potential.}

The vortices in the lattice Abelian Higgs model are defined as follows \cite{DeGrand:1980eq}.
Rewriting the Higgs field as $\phi=\rho e^{i\theta}$, we can reduce a redundant gauge degree of freedom by the transformation $\bar A_\mu(x) \equiv A_\mu(x) + \theta(x+\hmu)-\theta(x)$.
The reduced field strength
$\bar F_{\mu\nu}(x) \equiv \bar A_\mu(x)+\bar A_\nu(x+\hat\mu)-\bar A_\mu(x+\hat\nu)-\bar A_\nu(x)$
is equivalent to the sum of the circulations of the gauge and Higgs fields.
As $\bar A_\mu$ is defined in $(-\pi,\pi]$, $\bar F_{\mu\nu}$ takes the value in $(-4\pi,4\pi]$.
It can be decomposed into large discrete and small continuous parts.
For example, the reduced field strength in the $x$-$y$ plane is decomposed as
\begin{equation}
\label{eqnb}
 \bar F_{xy}(x) = 2\pi n(x) + B(x)
\end{equation}
with
\begin{equation}
 n(x)  \in \{-2,-1, 0, 1, 2\},
\qquad
 B(x) \in (-\pi,\pi].
\end{equation}
In the Higgs phase, $n$ corresponds to the winding number of the Higgs field and $B$ corresponds to a magnetic field in the $z$ direction.

We consider a static and straight vortex along the $z$ direction.
The constraint condition for the vortex is given by the product of the Kronecker delta,
\begin{equation}
\label{eqC}
C(\vec r) = \prod_{z,\tau} \delta_{n(x) \pm 1}
.
\end{equation}
The vector $\vec r$ is the 2-dimensional coordinate in the $x$-$y$ plane.
The double sign corresponds to the vortex $n = +1$ and the antivortex $n = -1$.
When the two vortices in Fig.~\ref{fig12} exist, the constrained path integral is
\begin{equation}
\label{eqZ2}
 Z(\vec r_1, \vec r_2) = \int D A D\phi D\phi^* \ e^{-S} C(\vec r_1) C(\vec r_2) 
.
\end{equation}
The expectation value of an observable $\mathcal{O}$ is defined by
\begin{equation}
 \langle \mathcal{O} \rangle = \frac{1}{Z(\vec r_1, \vec r_2)} \int DA D\phi D\phi^* \ \mathcal{O} \ e^{-S} C(\vec r_1) C(\vec r_2) 
.
\end{equation}
The expectation value is stochastically evaluated by the Monte Carlo method with probability $P \equiv e^{-S} C(\vec r_1) C(\vec r_2) $.
We implemented it to the Metropolis algorithm.
The algorithm is started with an initial configuration satisfying the constraint conditions.
Then the configuration is randomly updated.
If the constraint conditions are still satisfied, i.e., $P = e^{-S}$, the update is accepted with probability $e^{-S}$.
If the constraint conditions are not satisfied, i.e., $P=0$, the update is rejected.
These are repeated many times.

Let us introduce the energy density, which is defined by the temporal component of the energy-momentum tensor \cite{Yanagihara:2019foh},
\begin{equation}
\begin{split}
 \varepsilon(x) =& \ \mathcal{L}(x) - 2\beta \sum_{\mu} \{ 1 - \cos F_{\mu\tau}(x) \}
\\
&- 2 \left| \phi(x+\hat{\tau}) - e^{iA_\tau(x)} \phi(x) \right|^2
.
\end{split}
\end{equation}
The intervortex potential is given by the total energy per length
\begin{equation}
\label{eqV}
V(r) = \frac{1}{L_z} \int d^3x \langle \varepsilon(x) \rangle
\end{equation}
up to a constant.
The potential is a function of the intervortex separation $r = |\vec r_1 - \vec r_2|$.
The definition \eqref{eqV} goes to infinity in the infinite-volume limit because of the 3-dimensional volume integral, but such infinite-volume divergence is a $r$-independent constant.
The potential can be made finite by subtracting the constant $V(r=\infty)$.

We applied the above formulation to the Higgs phase to get the intervortex potential in superconductors.
Although we can naively apply it to the Coulomb and confinement phases, the interpretation of the results needs attention.
Even without the Higgs condensate, lattice gauge fields can form topological defects, the so-called monopole currents \cite{DeGrand:1980eq}.
Even in pure gauge theory (i.e., the XY model), the monopole currents exist.
The resultant potential should be interpreted as the potential of the monopole currents.
It would be interesting from the viewpoint of lattice gauge theory, as the monopole currents play important roles in the confinement phase \cite{Baig:1998ui,Wenzel:2005nd}.
It is however physically different from the potential of superconducting vortices, so we do not analyze it in this work.

\paragraph{Simulation results.}

We did the numerical simulations with the following condition:
Lattice volume is $L_x L_y L_z L_\tau = 12^4$ and boundary conditions are periodic.
The maximum intervortex distance is $L_x/2=6$.
We subtracted the constant $V(r=6)$ from $V(r)$.
We fixed $\beta = 1.5$ and $v=2.0$.
The critical coupling constant of the Coulomb-Higgs phase transition is $\lambda_{\rm C} \simeq 0.049$.
We show the results in the Higgs phase $\lambda > \lambda_{\rm C}$.

\begin{figure}[t]
\begin{center}
 \includegraphics[width=.49\textwidth]{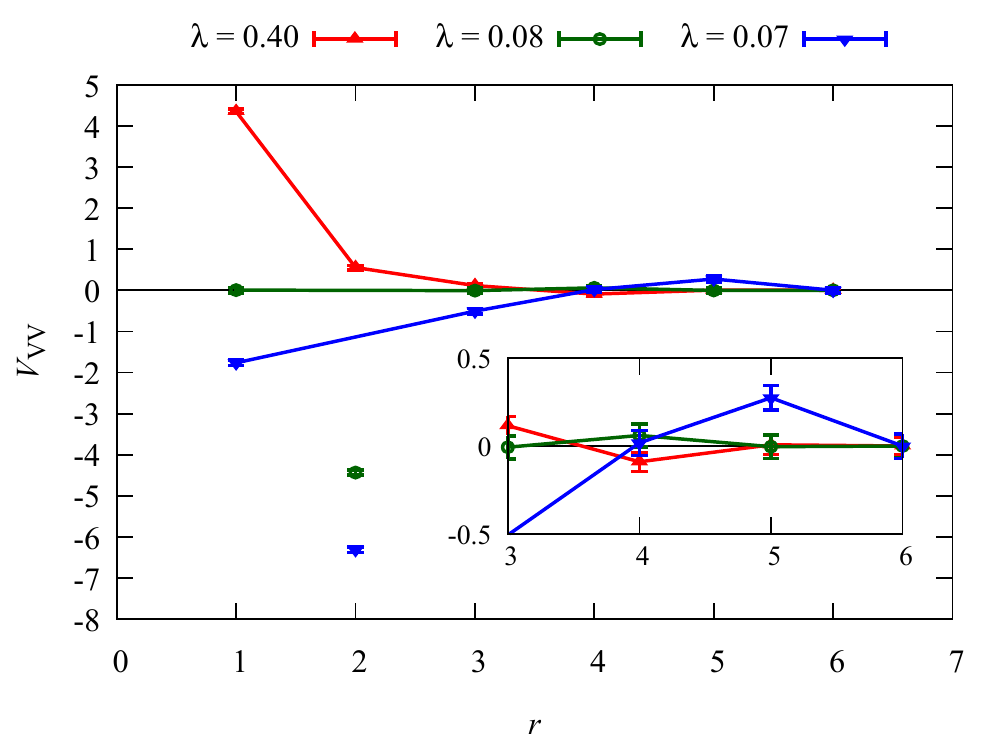}
\caption{
\label{figVV}
Vortex-vortex potential $V_{\rm VV}(r)$.
Colored solid lines are guides for the eye.
The data which are not connected by the lines are the ``topological transition points''.
}
\end{center}
\end{figure}

The vortex-vortex potential is shown in Fig.~\ref{figVV}.
At a small coupling constant $\lambda=0.07$, the potential is mainly attractive although weak repulsion is seen at long range $r = 5$-6.
This implies that the system is type I.
At a large coupling constant $\lambda=0.40$, the potential is repulsive, and therefore the system is type II.
Between these two cases, at $\lambda=0.08$, the potential is flat.
The two vortices are non-interacting with each other.
Such a non-interacting state is called the Bogomol'nyi-Prasad-Sommerfield (BPS) or critical vortex.
The appearance of the BPS vortex means the type-I-type-II transition.
Its critical coupling constant is estimated as $\lambda_{\rm H} \simeq 0.08$.
These results are consistent with the phase diagram in Fig.~\ref{fig12}.
We see a strange behavior at $\lambda=0.07$ and 0.08.
The potential suddenly changes at $r=2$.
We call it ``topological transition point'' for the reason explained below.

\begin{figure}[t]
\begin{center}
 \includegraphics[width=.49\textwidth]{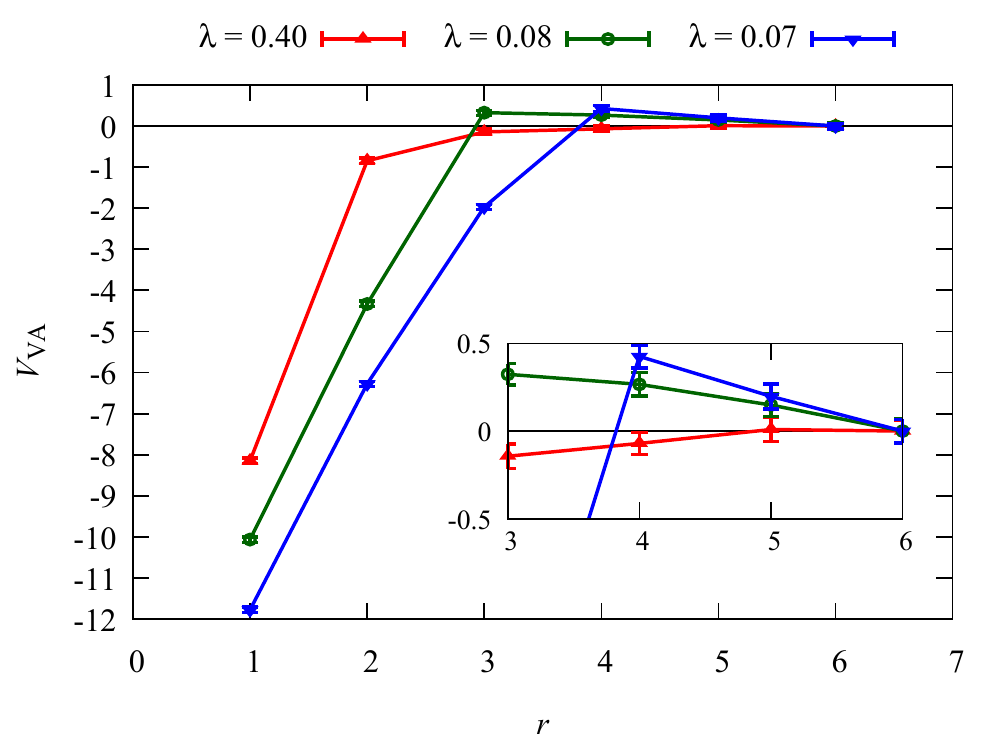}
\caption{
\label{figVA}
Vortex-antivortex potential $V_{\rm VA}(r)$.
Colored solid lines are guides for the eye.
}
\end{center}
\end{figure}

The vortex-antivortex potential is shown in Fig.~\ref{figVA}.
At short range, the potential is always attractive.
This is consistent with our physical intuition.
The zero-vortex and zero-antivortex state must be the lowest energy state.
The vortex and antivortex tend to merge and annihilate.
The potential is not flat at $\lambda = 0.08$, but it does not contradict the property of the BPS vortex.
While the BPS vortex restricts the vortex-vortex interaction to be zero, it does not restrict the vortex-antivortex interaction at all.
The vortex-antivortex potential is attractive at short range, and therefore cannot be completely flat.
At long range, the potential is sensitive to the coupling constant.
As shown in the inset of Fig.~\ref{figVA}, the potential is repulsive at $\lambda = 0.07$ (type-I) and attractive at $\lambda = 0.40$ (type-II).
This is opposite to the $\lambda$-dependence of the vortex-vortex potential.

\begin{figure}[t]
\begin{center}
 \includegraphics[width=.49\textwidth]{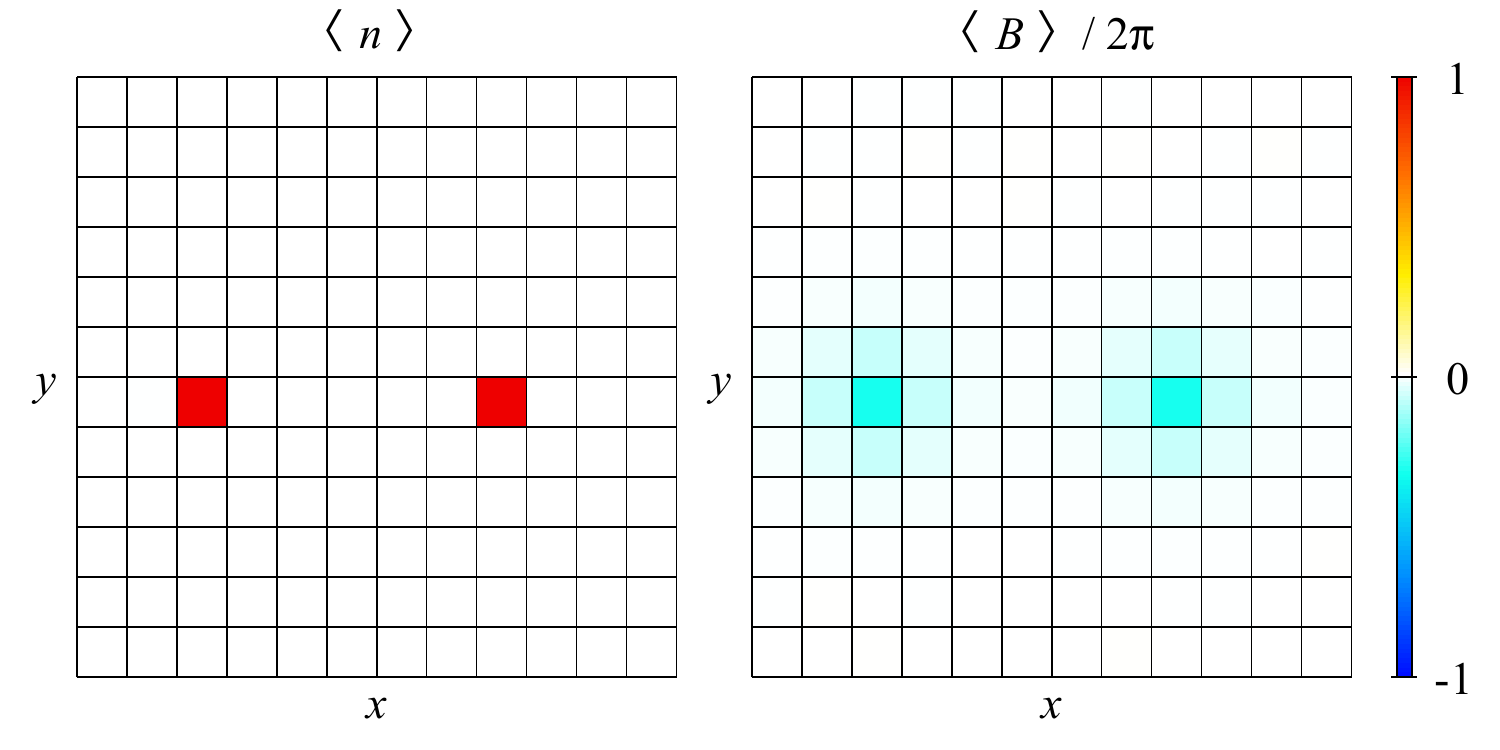}
\caption{
\label{fig3D}
Winding number $\langle n \rangle$ and magnetic field $\langle B \rangle$ for $V_{\rm VV}(r=6)$ with $\lambda=0.07$.
Statistical error bars are omitted.
}
\end{center}

\begin{center}
 \includegraphics[width=.49\textwidth]{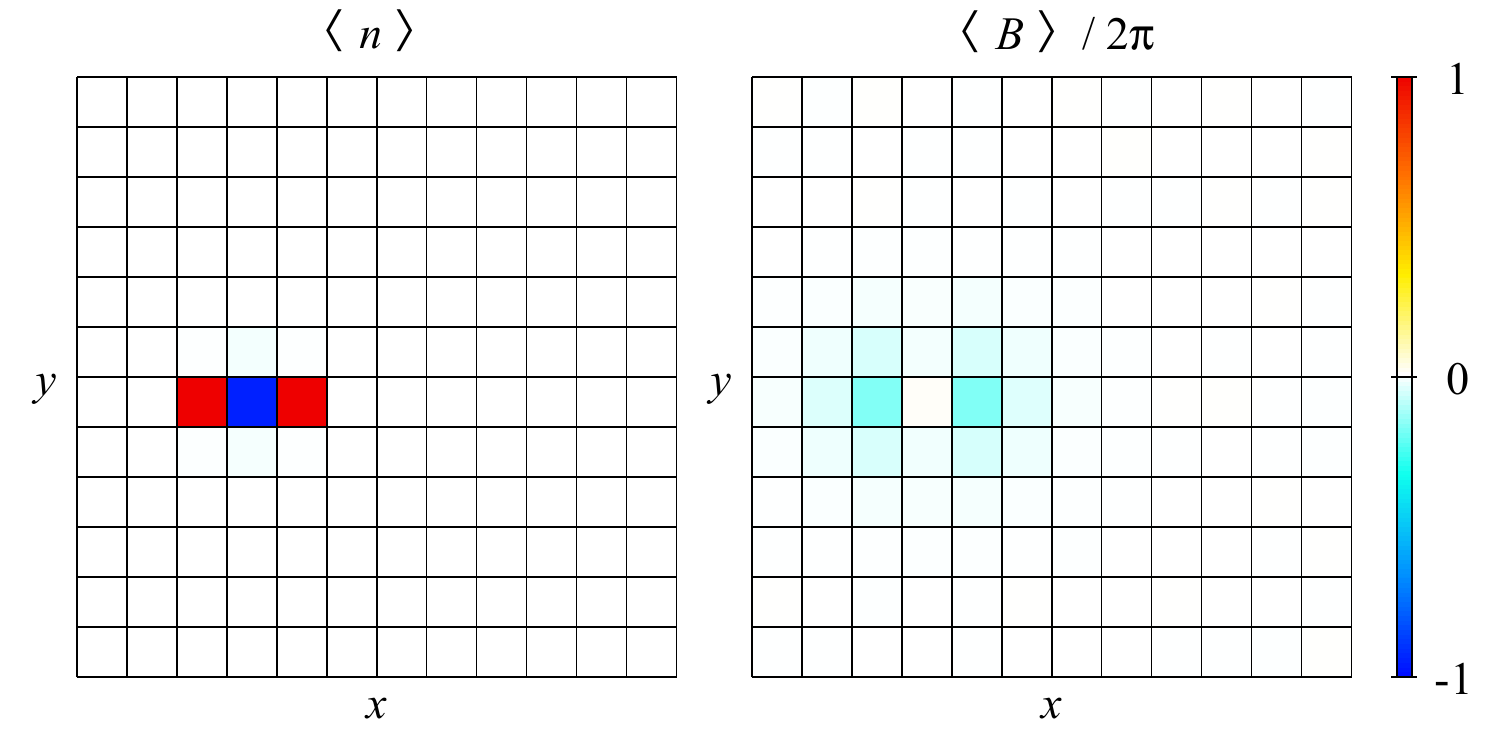}
\caption{
\label{fig3D2}
Same as Fig.~\ref{fig3D}, but for $V_{\rm VV}(r=2)$.
}
\end{center}
\end{figure}

What happens at the topological transition point $r=2$?
To answer this question, let us focus on the distribution of winding number $n$.
A typical case of $r \neq 2$ is shown in Fig.~\ref{fig3D}.
Vortices exit only at the positions where the constraint conditions are imposed.
The total winding number is $N \equiv \int dxdy \langle n \rangle= 2$.
The distribution of magnetic field $B$ is also shown in Fig.~\ref{fig3D}.
Negative magnetic fields are induced around the vortices because superconductors tend to expel magnetic fluxes.
At $r=2$, as shown in Fig.~\ref{fig3D2}, in addition to the two constrained vortices, one dynamical antivortex is induced at the midpoint of them.
The total winding number is $N = 1$.
Even though the creation of an antivortex costs the extra energy per length $E_{\rm A}$, the vortex-antivortex potential $V_{\rm VA}$ in Fig.~\ref{figVA} is strongly negative.
As a naive estimation, the $N=1$ state is favored when the inequality $E_{\rm A} + 2V_{\rm VA}(r/2)<0$ is satisfied.
We found that the topological transition point is sensitive to parameters:
(i) It does not appear at $\lambda = 0.40$.
The scalar self-coupling term favors condensate formation and disfavors antivortex creation.
The antivortex-creation energy $E_{\rm A}$ will be large, so the inequality is violated.
(ii) The $N=1$ state is favored at short range, but it is not seen at $r=1$ in Fig.~\ref{figVV}.
This would be due to lattice artifact.
Since the lattice has no room between two vortices at $r=1$, an antivortex must be created away from the midpoint.
The energy gain by the vortex-antivortex potential $V_{\rm VA}$ will be small, so the inequality is violated.

The physical relevance of the topological transition depends on what situation is considered.
In our definition of the potential, the winding numbers of the two constrained vortices are fixed but the total winding number is not fixed.
This allows the topological transition from $N=2$ to $N=1$.
On the other hand, if the total winding number should be fixed, the topological transition cannot occur.
We must introduce the constraint condition for $N$; it is actually straightforward.

\paragraph{Discussion.}

As discussed above, our results are qualitatively consistent with the classical picture.
We comment on a few differences from the previous classical analysis.

We observed the topological transition point.
In principle, we can observe it even in the classical analysis because the topological transition is not a quantum but classical effect.
In practice, however, its numerical treatment is difficult.
In the standard classical analysis, the winding number distribution is a priori assumed or field configuration is obtained by smooth deformation.
The topological transition is forbidden in such analysis. 

The obtained potentials are quite different from the potentials in the linearized approximation \cite{Speight:1996px}.
The linearized potential is proportional to the winding number $n$ by construction.
The vortex-vortex potential and the vortex-antivortex potential are antisymmetric.
In Figs.~\ref{figVV} and \ref{figVA}, the antisymmetry is apparently violated even at long range.

The critical coupling constant of the type-I-type-II transition is $\lambda_{\rm H} \simeq 0.08$.
In the classical continuous theory, the transition happens at the critical Ginzburg-Landau parameter $\kappa \equiv \sqrt{\lambda}/e = 1/\sqrt{2}$.
This implies the critical coupling constant $\lambda_{\rm H}=1/(2\beta)=1/3$.
The deviation will be due to both of quantum effect and discretization artifact.

\paragraph{Concluding remarks.}

We obtained the intervortex potential by the first-principles calculation of the quantum Monte Carlo method.
The calculation is free from uncontrollable approximation or assumption.
Although the continuum and infinite-volume limits were not extrapolated, the extrapolation will be systematic; it is a future work.
Another future work is to introduce finite mass of vortices.
The constraint condition \eqref{eqC} generates a static, i.e, infinitely heavy, vortex, so we can access only the potential of static vortices.
It is dual to the potential of static charges, which can be obtained by the conventional calculation of the Wilson lines.
Nowadays, the potential of finite-mass particles is intensively studied by the state-of-the-art technique in lattice gauge theory \cite{Aoki:2012tk}.
If our formulation can be combined with such a technique, we can study the potential of finite-mass vortices, which is closer to reality.
The finite-mass vortices can form bound states.
We can also discuss the bound-state formation by another technique in lattice gauge theory \cite{Luscher:1990ux}.

The application of the above formulation is not limited to the Abelian Higgs model.
We can apply it to the microscopic theory with electrons, instead of the Higgs field.
As another application, we can study the color-dependent potentials of non-Abelian vortices, which are relevant for the physics of neutron stars \cite{Eto:2013hoa}, by the lattice gauge simulation of non-Abelian Higgs models \cite{Yamamoto:2018vgg}.
It will be also applicable to the theories with higher- or lower-dimensional topological defects, such as monopoles, domain walls, etc.

\begin{acknowledgements}
The author greatly thank Tomoya Hayata for the cooperation in early stage.
The author was supported by JSPS KAKENHI Grant No.~19K03841.   
The numerical simulation was carried out on SX-ACE in Osaka University.
\end{acknowledgements}

\bibliography{./paper}

\end{document}